# Determining a Quantum Theory of the Infinite-Component Majorana Field

L. Nanni†
University of Ferrara
44100-Ferrara, Italy
†E-mail: luca.nanni68@gmail.com

**Abstract**: In this paper, the quantum theory of the infinite-component Majorana field for the fermionic tower is formulated. This study proves that the energy states with increasing spin are simply composite systems made by a bradyon and antitachyons with half-integer spin. The quantum field describing these *exotic states* is obtained by the infinite sum of four-spinor operators, which each operator depends on the spin and the rest mass of the bradyon in its fundamental state. The interaction between bradyon-tachyon, tachyon-tachyon and tachyon-luxon has also been considered and included in the total Lagrangian. The obtained theory is consistent with the CPT invariance and the spin-statistics theorem and could explain the existence of new forms of matter not predictable within the standard model.

## 1   Introduction

The formulation of the quantum field theory based on the finite-dimensional representation of the Lorentz group led to the standard model (SM) and coherently explained most of the experimental results [1–4]. However, the representation is neither complete nor unitary [5, 6]. This gap can be filled through an infinite-dimensional representation, where boosts and rotations take the form of infinite matrices. In principle, a theory whose equations are covariant with respect to the infinite-dimensional representations could predict new particles or new structures of matter not contemplated in the picture of the SM. Although with different aims, Majorana formulated in 1932 a relativistic equation for particle with arbitrary spin [7], which it is a *universal* equation that describes the physical nature of bosons, fermions and luxons that depends on the considered spin and mass. In other words, all spins are simultaneously representations of the inhomogeneous Lorentz groups obtained while only considering the spacetime symmetries [8]. The Majorana equation has two possible applications depending on the interpretation given to the wave function: particles with arbitrary spin or composite systems [8–11]. Therefore, this equation may be particularly useful in studying the structure of nuclear systems and their prospective high energy *exotic* states. However, the solution of the Majorana equation leads to results that contrast with physical reality (particle with only positive frequency, mass spectrum that asymptotically decreases with spin increasing, spacelike solutions) [12]. Moreover, the attempt to quantise the infinite-component Majorana field violates CPT invariance and is inconsistent with the spin-statistic theorem. In addition, explaining the existence of tachyonic solutions that are incoherent with the *classical* theory of relativity is difficult [12–14]. Sudarshan managed these difficulties while separately investigating the three classes of particles predicted by the Majorana equation: slower than light particles, luxons and faster than light particles [15]. The method used by Sudarshan, therefore, is indirect and is based on the decomposition of the Majorana spinor, i.e. its bradyonic, tachyonic and luminal components. However, a direct method to quantise the Majorana field in a coherent manner with the fundamental theorems of modern quantum mechanics does not yet exist. Moreover, little has been done to apply Majorana field theory to composite (multimass) systems [9].

This study investigates the physical nature of the energy states with increasing spins that form the Majorana bradyonic tower, but this study is limited to only investigating fermions. These states are also proven to be simply *exotic* composite systems made by a bradyon and antitachyons, which all these components with half-integer spin. In this sense, the bradyonic tower includes all the possible solutions of the infinite-component equation without needing to face the quantisation process for each type of particles (bradyon, tachyon, luxon). However, once again, the field theory is formulated in an indirect way as the sum of the (local) Dirac field [16] with positive frequency and the tachyonic (local) field with negative frequency [17]. These two fields are connected to each other by a Lorentz superluminal transformation [18-19], which also represents the bradyon-tachyon interaction mechanism. The Majorana field is, therefore, the infinite sum of four-spinor operators which depend on the spin $J$. In forming this theory, the tachyon-tachyon and tachyon-luxon interactions are considered. Because of the lack of experimental data, these interactions are introduced in the Lagrangian, making use of some speculative theories available in the scientific literature [20,21]. Therefore, the total Lagrangian is the following:

$$\mathcal{L}_M = \mathcal{L}_{D_+} + \mathcal{L}_{t_-} + \mathcal{L}_{int.} \qquad (1)$$

where $\mathcal{L}_M$ is the Lagrangian of Majorana field, $\mathcal{L}_{D_+}$ is the Lagrangian of the Dirac field with positive frequencies, $\mathcal{L}_{t_-}$ is the Lagrangian of the tachyonic field with negative frequencies and $\mathcal{L}_{int.}$ is the Lagrangian of the interactions between the local fields. Overall, the obtained theory is consistent with the CPT invariance and with the spin-statistics theorem.

## 2 The Majorana Bradyonic Tower as Composite Systems

The solution of the Majorana equation for arbitrary spin leads to a discrete mass spectrum:

$$m(J) = \frac{m_0}{\left(\frac{1}{2}+J\right)} \qquad (2)$$

where $m_0$ is the rest mass of the particle in the fundamental state. In addition, this study only considers fermions with half-integer spin, i.e. $J = \frac{1}{2}, \frac{3}{2}, \frac{5}{2}, \ldots$

The occupation probability of a Majorana state with spin $J$ is the following [22]:

$$p(J) = \left[\beta^n - \beta^{(n+1)}\right]^{1/2} \qquad (3)$$

where $n = (1/2 + J)$ and $\beta$ is the relativistic factor. Using the energy-momentum relation, the energy difference between whatever state with spin $J$ and the fundamental state is the following:

$$E^2(J) - E^2(J_0) = [p^2(J) - p^2(J_0)]c^2 - \frac{(J+1/2)^2-1}{(J+1/2)^2} m_0^2 c^4 \qquad (4)$$

Equation **(4)** is the typical form of the energy-momentum relation for a tachyon with imaginary mass:

$$\mu(J) = i\frac{\sqrt{(J+1/2)^2-1}}{(J+1/2)} m_0 \qquad (5)$$

Equation **(5)** suggests that the Majorana states with decreasing mass are the results of the interactions between the ½-spin bradyon quantum field and the ½-spin tachyon field. To this purpose, the Majorana equation predicts a discrete bradyonic mass spectrum with only positive frequency and a continuous imaginary mass spectrum with both positive and negative frequencies [7]. Therefore, Majorana states with spin $J > J_0$ may be considered as composite systems made by a 1/2-spin bradyon of rest mass $m_0$ and 1/2-spin tachyons with imaginary mass. Starting from this assumption, a field theory consistent with the spin-statistic theorem and whose Hamiltonian operator is upper bound is established.

## 3 The Dirac Field with Positive Energies

The Dirac field is a four-dimension spinor that transforms from a given reference frame to another under the action of the symmetry elements of the Lorentz group in its four-dimensional representation [1]. The Lagrangian density of the Dirac group is the following:

$$\mathcal{L}_D = \bar{\psi}_D(i\gamma^\mu \partial_\mu - m_0)\psi_D \tag{6}$$

where $\gamma^\mu$ are the four Dirac matrices (μ=0,…,3) and the conjugate field is the following:

$$\bar{\psi}_D = \psi_D^\dagger \gamma^0 \tag{7}$$

The Dirac field current is the following:

$$J^\mu(x) = \bar{\psi}_D \gamma^\mu \psi_D \tag{8}$$

whose divergence is identically zero:

$$\partial_\mu J^\mu(x) = (\partial_\mu \bar{\psi}_D)\gamma^\mu \psi_D + \bar{\psi}_D \gamma^\mu \partial_\mu \psi_D = im_0 \bar{\psi}_D \psi_D + \bar{\psi}_D(-im_0 \psi_D) = 0 \tag{9}$$

Therefore, the Dirac current is a conserved quantity. In addition, the Dirac field is quantised by the anticommutation rule:

$$\begin{cases} \{\psi_a, \psi_b\} = \delta^{(3)}(x - y)\delta_{ab} \\ \{\psi_a, \psi_b\} = \{\psi_a^\dagger, \psi_b^\dagger\} = 0 \end{cases} \tag{10}$$

As explained in the previous section, the solutions with positive energy are desired, which are the same solutions as the Majorana equation for a particle with 1/2-spin [7]. To do this, $\gamma^0 = \mathbb{1}$ must be set so that relation **(7)** becomes the following:

$$\bar{\psi}_{D_+} = \psi_{D_+}^\dagger \tag{11}$$

and equation **(6)** becomes the following:

$$\mathcal{L}_D = \psi_{D_+}^\dagger (i\gamma^\mu \partial_\mu - m_0)\psi_{D_+} \tag{12}$$

The field operators are obtained solving the Lagrangian equations, and their Fourier expansions in terms of creator and annihilator operators are as follows [1]:

$$\begin{cases} \bar{\psi}_{D_+} = \int \frac{d^3 p}{(2\pi)^3} \frac{1}{\sqrt{2E_{D_+}}} \sum_s (a_p^s)^\dagger \bar{u}^s(p) \exp\{-i(\mathbf{kx} - \omega t)\} \\ \psi_{D_+} = \int \frac{d^3 p}{(2\pi)^3} \frac{1}{\sqrt{2E_{D_+}}} \sum_s a_p^s u^s(p) \exp\{i(\mathbf{kx} - \omega t)\} \end{cases} \tag{13}$$

where $E_{D_+} = \sqrt{p^2 c^2 + m_0^2 c^4}$, $\omega = E_{D_+}/\hbar$ and $\mathbf{k} = \mathbf{p}/\hbar$, while $u^s(p)$ is the spinor function. The creator and annihilator operators must satisfy the anticommutation relation:

$$\{a_p^r, (a_q^s)^\dagger\} = (2\pi)^3 \delta^{(3)}(\mathbf{p} - \mathbf{q})\delta_{rs} \tag{14}$$

The current **(8)** becomes the following:

$$J^\mu(x) = \psi_{D_+}^\dagger \gamma^\mu \psi_{D_+} \tag{15}$$

whose gradient is always zero:

$$\partial_\mu J^\mu(x) = (\partial_\mu \psi_{D_+}^\dagger)\gamma^\mu \psi_{D_+} + \psi_{D_+}^\dagger \gamma^\mu \partial_\mu \psi_{D_+} = im_0 \psi_{D_+}^\dagger \psi_{D_+} + \bar{\psi}_D(-im_0 \psi_{D_+}) = 0 \tag{16}$$

Equation **(16)** shows that $\psi_{D_+}^\dagger = (\psi_{D_+})^{-1}$, which agrees with the Majorana equation [7] for $j_0 = 1/2$. Therefore, the field current is conserved for the Dirac field with positive energies. Moreover, since $\gamma^0 = \mathbb{1}$, the explicit form of the current is the following:

$$J^\mu(x) = \psi_{D_+}^\dagger \psi_{D_+} + \psi_{D_+}^\dagger \gamma^\mu \psi_{D_+} = 1 + \psi_{D_+}^\dagger \gamma^\mu \psi_{D_+} \quad \gamma = 1,2,3 \tag{17}$$

Equation **(17)** shows that the current always has a positive time component, which confirms that the particle is a bradyon.

Having limited the Dirac field to the only positive frequencies, the Hamiltonian density becomes the following:

$$\mathcal{H} = \int \frac{d^3 p}{(2\pi)^3} E_{D_+}(p) \sum_s (a_p^s)^\dagger a_p^s \tag{18}$$

Equation **(18)** is not upper bound and, as expected, is similar to the Hamiltonian obtained when the Majorana field is directly quantised [12]. However, for our purpose, the Dirac field with positive energies is only a *tool* needed to coherently quantise the infinite-component field. Nonetheless, the *trouble* produced from equation **(18)** is overcome when in the desired theory the tachyonic field is introduced.

## 4 The Tachyonic Field with Negative Energies

The 1/2–spin tachyonic field theory is developed by limiting the Lemke equation [17] to solutions with negative energies. The Lagrangian density is the following:

$$\mathcal{L}_{L_-} = \bar{\psi}_{L_-}\left(i\gamma^\mu \partial_\mu + im\right)\psi_{L_-} \tag{19}$$

where $\gamma^\mu$ are the Dirac matrices with $\gamma^0 = \mathbb{1}$. In this theory, the mass $m$ must satisfy equation **(5)**; thus, the Lemke Lagrangian may be rewritten as the following:

$$\mathcal{L}_{L_-} = \bar{\psi}_{L_-}\left(i\gamma^\mu \partial_\mu + i\frac{\sqrt{(J+1/2)^2-1}}{(J+1/2)} m_0\right)\psi_{L_-} \tag{20}$$

whose negative energies are the following:

$$E_{L_-} = -\sqrt{[p^2(J) - p^2(J_0)]c^2 - \frac{(J+1/2)^2-1}{(J+1/2)^2} m_0^2 c^4} \tag{21}$$

Equation **(21)** has physical meaning only if the following constraint is verified:

$$[p^2(J) - p^2(J_0)] \geq \frac{(J+1/2)^2-1}{(J+1/2)^2} m_0^2 c^2 \tag{22}$$

Therefore, once the impulse of the starting bradyon with $J_0 = 1/2$ is defined, the antitachyon interacting with it must have an impulse such that the constraint **(22)** is fulfilled. This means that the Majorana state with spin $j$ must be the following:

$$p^2(J) \geq p^2(J_0) + \frac{(J+1/2)^2-1}{(J+1/2)^2} m_0^2 c^2 \tag{23}$$

At the limit $j \to \infty$, the squared impulse becomes equal to $(p^2(J_0) + m_0^2 c^2)$, i.e. by increasing the spin the impulse of the *exotic* state reaches a minimum value different from zero. Furthermore, relativistic kinematics shows that the velocity of the bradyon in the *exotic* state must be greater than $c/\sqrt{2}$. Equation **(23)** also shows that when the velocity of the bradyon in the *exotic* state with spin $J$ approaches the speed of light, the impulse $p(J_0)$ tends to $m_0 c$ and the impulse $p(J)$ of the composite particle tends to zero. Therefore, to this kinematic limit, the antitachyon energy tends to zero, thus being upper bound.

The tachyonic field with negative energy as Fourier expansions of creator annihilator operators is the following:

$$\begin{cases} \bar{\psi}_{L_-} = \int \frac{d^3 p}{(2\pi)^3} \sum_{s,j} \frac{1}{\sqrt{2E_p}} \left(t_p^{s,j}\right)^\dagger \bar{w}^{s,j}(\boldsymbol{p}) \exp\{-i(\boldsymbol{kx} - \omega t)\} \\ \psi_{L_-} = \int \frac{d^3 p}{(2\pi)^3} \sum_{s,j} \frac{1}{\sqrt{2E_p}} t_p^{s,j} w^{s,j}(\boldsymbol{p}) \exp\{i(\boldsymbol{kx} - \omega t)\} \end{cases} \tag{24}$$

In this case, the sum runs also on the spin index, and the quantities $\boldsymbol{k}$ and $\omega$ depend on it:

$$\begin{cases} \omega^2(J) = \frac{1}{\hbar^2}\left[\left(p^2(J)c^2 - \frac{m_0^2 c^4}{(J+1/2)^2}\right) - (p^2(J_0)c^2 - m_0^2 c^4)\right] \\ k^2(J) = \frac{1}{\hbar^2}\left(p^2(J) - p^2(J_0)\right) \end{cases} \tag{25}$$

Therefore, the exponential functions in the Fourier expansions **(24)** may be written as follows:

$$\exp\{\pm i(\boldsymbol{kx} - \omega t)\} = \exp\{\pm i(\boldsymbol{k}(j)\boldsymbol{x} - \omega(j)t)\}\exp\{\pm i(\boldsymbol{k}(j_0)\boldsymbol{x} - \omega(j_0)t)\} \tag{26}$$

The operator $\left(t_p^{s,j}\right)^\dagger$ creates a tachyon with an impulse that, in accordance with equation **(23)**, depends on the momentum of the 1/2-spin bradyon. This means that the creator and annihilator tachyonic operators implicitly depend on those bradyonic with positive energies.

Since $\gamma^0 = \mathbb{1}$, the conjugate tachyonic field is equal to its adjoint, and the scalar product between the two fields $\bar{\psi}_{L_-}$ and $\psi_{L_-}$ is zero [17]:

$$\bar{\psi}_{L_-}\psi_{L_-} = \psi_{L_-}^\dagger \gamma^0 \psi_{L_-} = 0 \tag{27}$$

By the orthogonality property **(27),** it follows that the tachyonic current is the following:
$$J^\mu_{L_-}(x) = \psi^\dagger_{L_-}\psi_{L_-} + \psi^\dagger_{L_-}\gamma^\mu\psi_{L_-} = \psi^\dagger_{L_-}\gamma^\mu\psi_{L_-} \quad \gamma = 1,2,3 \tag{28}$$
As expected, the tachyonic current does not have the time coordinate but only the space coordinates.

The next step should be the quantisation of the tachyonic field, but, the following section proves that the tachyonic field is connected with the Dirac field by SLT symmetry and that the anticommutation relations fulfilling the Dirac field also hold for the tachyonic one. In other words, the quantisation of the Dirac field with positive energies also implies quantisation of the tachyonic field with negative energies.

## 5  Algebraic Connection Between the Bradyonic and Tachyonic Fields

The infinite-component Majorana equation provides both bradyonic and tachyonic solutions [7]. This suggests that the two fields should coexist and that one should be the algebraic connection of the other. This is equivalent to claiming that a symmetry transformation exists between the bradyonic operators $(a^s_p)^\dagger$ and $a^s_p$ with the tachyonic ones $(t^{s,j}_p)^\dagger$ and $t^{s,j}_p$ and that a similar transformation exists between the spinors $u^s(p)$ and $w^{s,j}(p)$. In other words, an invertible transformation that changes a timelike solution in a spacelike solution is determined. This symmetry transformation is the *interaction force* which holds together the *exotic* state formed by a 1/2-spin bradyon and a given number of 1/2-spin antitachyons to be compatible with the total spin $J$. In agreement with Recami [18,19], the desired SLT transformation is antiunitary:
$$\Lambda^T \Lambda = -\mathbb{1} \tag{29}$$
The transformation depends on the spin J such that acting on the bradyonic operator $a^s_p$ gives the tachyonic operator $t^{s,j}_p$. This transformation occurs through the similarity transformation:
$$(\Lambda_J)^{-1} a^s_p \Lambda_J = t^{s,j}_p \tag{30}$$
Using the properties **(29),** equation **(30)** may be rewritten as the following:
$$(\Lambda_J)^{-1} a^s_p \Lambda_J = -\Lambda_J^{-1} a^s_p \Lambda_J (\Lambda_J^{-1}\Lambda_J) = -t^{s,j}_p \Lambda_J^{-1}\Lambda_J \tag{31}$$
Furthermore, applying equation **(31)** to the spinor gives the following:
$$[-\Lambda_J^{-1} a^s_p \Lambda_J(\Lambda_J^{-1}\Lambda_J)]w^{s,j}(p) = -\Lambda_J^{-1} a^s_p (\Lambda_J\Lambda_J^{-1})\Lambda_J w^{s,j}(p) =$$
$$= -\Lambda_J^{-1} a^s_p \left(\Lambda_J w^{s,j}(p)\right) = -\Lambda_J^{-1} a^s_p u^s(p) \tag{32}$$
Equation **(32)** is the algebraic transformation, linking the bradyonic state with rest mass $m_0$ and the tachyonic states with mass $m(J)$. The following equations then generalise the transformations:
$$\begin{cases} t^{s,j}_p w^{s,j}(p) = -\Lambda_J^{-1} a^s_p u^s(p) \\ (t^{s,j}_p)^\dagger \overline{w}^{s,j}(p) = -\Lambda_J^{-1} (a^s_p)^\dagger \bar{u}^s(p) \end{cases} \tag{33}$$
The antitachyon field operators become the following:
$$\begin{cases} \bar\psi_{L_-} = -\int \frac{d^3p}{(2\pi)^3} \sum_{s,j} \frac{1}{\sqrt{2E_p}} \Lambda_J^{-1} (a^s_p)^\dagger \bar{u}^s(p) \exp\{-i(kx - \omega t)\} \\ \psi_{L_-} = -\int \frac{d^3p}{(2\pi)^3} \sum_{s,j} \frac{1}{\sqrt{2E_p}} \Lambda_J^{-1} a^s_p u^s(p) \exp\{i(kx - \omega t)\} \end{cases} \tag{34}$$
Since the bradyonic operators $a^s_p$ and $(a^s_p)^\dagger$ satisfy the anticommutation relations **(14)**, the tachyonic operators comply with the same relations, as follows:
$$\{t^{s,j}_p, (t^{s,j}_p)^\dagger\} = \Lambda_J^{-1} a^s_p \Lambda_J \Lambda_J^{-1} (a^s_p)^\dagger \Lambda_J + \Lambda_J^{-1} (a^s_p)^\dagger \Lambda_J \Lambda_J^{-1} a^s_p \Lambda_J =$$
$$= -\Lambda_J^{-1} a^s_p (a^s_p)^\dagger \Lambda_J - \Lambda_J^{-1} (a^s_p)^\dagger a^s_p \Lambda_J = -\Lambda_J^{-1} \{a^r_p, (a^s_q)^\dagger\} \Lambda_J =$$
$$= -(2\pi)^3 \delta^{(3)}(p - p')\delta_{rs} \tag{35}$$

Relation **(35)** proves that the tachyonic field is quantised according to the anticommutation rules. Following the same approach, the anticommutation relation between the tachyonic fields shown in equation **(24)** is proven by the following relation:

$$\{\psi_{L_-}, \psi_{L_-}^\dagger\} = \delta^{(3)}(\boldsymbol{p} - \boldsymbol{p}') \tag{36}$$

## 6 The Infinite-component Majorana Field

In section 4, the relationship between the impulses of the *exotic state* with spin $J$ and of the fundamental state was obtained and shown in equation **(23)**. Now the following constraint is established:

$$\sqrt{p^2(J_0)c^2 + m_0^2 c^4} = \sqrt{p^2(J)c^2 - \frac{m_0^2 c^4}{(J+1/2)^2}} \tag{37}$$

from which the following equation is obtained:

$$p^2(J) = p^2(J_0) + \frac{(J+1/2)^2 + 1}{(J+1/2)^2} m_0^2 c^2 \tag{38}$$

Therefore, the energy of the *exotic state* ranges within $\left[0, \sqrt{p^2(J_0)c^2 + m_0^2 c^4}\right]$, and the energy of the tachyonic component contributing to the *exotic state* is the following:

$$E_t = \varepsilon(J) E(J_0) \quad 0 \le \varepsilon(J) \le 1 \tag{39}$$

The phases of the exponential functions appearing in the Fourier expansion of the quantum fields are the following:

$$exp\{\pm i(\boldsymbol{k}_t \boldsymbol{x} - \omega_t t)\} = exp\{\pm \varepsilon(J) i(\boldsymbol{k}_0 \boldsymbol{x} - \omega_0 t)\} = g(\varepsilon(J)) exp\{\pm i(\boldsymbol{k}_0 \boldsymbol{x} - \omega_0 t)\} \tag{40}$$

where the wave vector and the pulsation of the bradyon in the fundamental state are denoted by $\boldsymbol{k}_0$ and $\omega_0$, respectively. $g(\varepsilon(J))$ is a continuous function to be found. Therefore, the antitachyon field operators may be rewritten as the following:

$$\begin{cases} \bar\psi_{L_-} = -\sum_J \frac{g(\varepsilon(J))}{\sqrt{\varepsilon(J)}} \Lambda_J^{-1} \bar\psi_{D_+} \\ \psi_{L_-} = -\sum_J \frac{g(\varepsilon(J))}{\sqrt{\varepsilon(J)}} \Lambda_J^{-1} \psi_+ \end{cases} \tag{41}$$

By **(41)**, the Lagrangian **(19)** may be written as the following:

$$\mathcal{L}_{L_-} = \sum_J \frac{g(\varepsilon(J))}{\sqrt{\varepsilon(J)}} \left[ \Lambda_J^{-1} \bar\psi_{D_+} (i\gamma^\mu \partial_\mu + im) \Lambda_J^{-1} \psi_{D_+} \right] \tag{42}$$

As a result, the tachyonic mass is the following:

$$\mu(J) = i \frac{\sqrt{(J+1/2)^2 - 1}}{(J+1/2)} m_0 = \eta(J) m_0 \tag{43}$$

When $\eta(J)$ ranges between $[0,1]$, the Lagrangian **(42)** becomes the following:

$$\mathcal{L}_{L_-} = \sum_J \frac{g(\varepsilon(J))}{\sqrt{\varepsilon(J)}} \left[ \Lambda_J^{-1} \bar\psi_{D_+} (i\gamma^\mu \partial_\mu - \eta(J) m_0) \Lambda_J^{-1} \psi_{D_+} \right] \tag{44}$$

The Lagrangian of the Majorana *exotic state* with spin $J$ can then be presented as the following:

$$\mathcal{L}_M(J) = \sum_J \frac{g(\varepsilon(J))}{\sqrt{\varepsilon(J)}} \left[ \Lambda_J^{-1} \bar\psi (i\gamma^\mu \partial_\mu) \Lambda_J^{-1} \psi - \eta(J) m_0 \Lambda_J^{-1} \bar\psi \Lambda_J^{-1} \psi \right] + \left[ \bar\psi (i\gamma^\mu \partial_\mu) \psi - m_0 \bar\psi \psi \right] \tag{45}$$

To simplify the notation, the subscripts distinguishing the bradyonic and tachyonic fields have been eliminated. Thus, the Lagrangian **(45)** must be equivalent to:

$$\mathcal{L}_M = \bar\psi_M (i\gamma_M^\mu \partial_\mu - m_0) \psi_M \tag{46}$$

The infinite-component Majorana field is denoted by $\psi_M$, while $\gamma_M^\mu$ represents the infinite matrices which develop in block, increasing the value of the spin:

$$\gamma_M^\mu = \begin{pmatrix} \sigma^\mu(1/2) & 0 & \cdots \\ 0 & \sigma^\mu(3/2) & \cdots \\ \vdots & \vdots & \ddots \end{pmatrix} \tag{47}$$

where $\sigma^\mu(n/2)$ is the spin matrix for the particle with spin $n/2$. Thus, by the equivalence of the Lagrangians **(45)** and **(46)**, the following is obtained:

$$\begin{cases} \bar{\psi}_M(i\gamma^\mu_M \partial_\mu)\psi_M \approx \left\{\sum_J \frac{g(\varepsilon(J))}{\sqrt{\varepsilon(J)}}\left[\Lambda_J^{-1}\bar{\psi}(i\gamma^\mu \partial_\mu)\Lambda_J^{-1}\psi\right] + \left[\bar{\psi}(i\gamma^\mu\partial_\mu)\psi\right]\right\} \\ m_0\bar{\psi}_M\psi_M \approx m_0\left\{\bar{\psi}\psi + \sum_J \eta(J)\Lambda_J^{-1}\bar{\psi}\Lambda_J^{-1}\psi\right\} \end{cases} \quad (48)$$

Since the numerical coefficients $\varepsilon(J)$ and $\eta(J)$ vary within the same range, they can be set to be equal. Moreover, since $\bar{\psi} = \psi^\dagger$, and the tachyonic fields $\psi_{L_-}$ and $\psi_{L_-}^\dagger$ are orthogonal, the second equation in **(48)** becomes the following:

$$\bar{\psi}_M\psi_M \approx \psi^\dagger\psi \quad (49)$$

This result was expected because the Majorana theory reduces for positive frequencies to that of the Dirac theory if $J = 1/2$. All other infinite components of the Majorana field are such that their product is zero, ensuring that the time-component of the current is only that associated to the bradyon. The space-components, instead, are the results of the contribution of both the bradyon and tachyon. Therefore, the SLT matrices $\Lambda_J^{-1}$ transform the Dirac fields $\psi$ and $\psi^\dagger$, making them orthogonal. Therefore, the Majorana field may be explicitly written as the following:

$$\psi_M = \begin{pmatrix} \psi \\ \sqrt{g(\varepsilon(3/2))}\Lambda_{3/2}^{-1}\psi \\ \sqrt{g(\varepsilon(5/2))}\Lambda_{5/2}^{-1}\psi \\ \vdots \end{pmatrix} \quad (50)$$

while the adjoining field is the following:

$$\psi_M^\dagger = \left(\psi^\dagger, -\sqrt{g(\varepsilon(3/2))}\Lambda_{3/2}^{-1}\psi^\dagger, -\sqrt{g(\varepsilon(5/2))}\Lambda_{5/2}^{-1}\psi^\dagger, \ldots\right) \quad (51)$$

Equation **(51)** shows that $(\Lambda_J^{-1})^\dagger = \Lambda_J^{-1}$, i.e. the STL matrices are antiorthogonal, which is expected since they are also antiunitary. Using the obtained results, the Majorana Lagrangian may be written as the following:

$$\mathcal{L}_M = \psi^\dagger\left\{i\left[\sum_J \frac{g(\varepsilon(J))}{\sqrt{\varepsilon(J)}}(\Lambda_J^{-1})^T\gamma^\mu\Lambda_J^{-1}\right]\partial_\mu - m_0\right\}\psi \quad (52)$$

Equation **(52)** shows that the action of the infinite Majorana matrices is equivalent to that of the infinite sum of Dirac matrices transformed by SLT transformations:

$$\gamma_M^\mu = \left[\sum_J \frac{g(\varepsilon(J))}{\sqrt{\varepsilon(J)}}(\Lambda_J^{-1})^T\gamma^\mu\Lambda_J^{-1}\right] \quad (53)$$

where the coefficient $\varepsilon(J)$ represents the expansion of the bradyonic tower with discrete mass spectrum, while the matrices $\Lambda_J$ represent the expansion of the tachyonic counterpart with a continuous mass spectrum. This is a relevant result, since an infinite matrix may be easily obtained using only finite 4x4 well-known matrices.

This theoretical approach also solves the problem affecting the Majorana Hamiltonian [12]; in fact, in terms of creator and annihilator operators, the Hamiltonian is the following:

$$\mathcal{H} = \int \frac{d^3\mathbf{p}}{(2\pi)^3}\sum_{s,J} E_\mathbf{p}(J)\left[(a_\mathbf{p}^s)^\dagger a_\mathbf{p}^s + (t_\mathbf{p}^{s,j})^\dagger t_\mathbf{p}^{s,j}\right] =$$

$$= \int \frac{d^3\mathbf{p}}{(2\pi)^3}\sum_{s,J} E_\mathbf{p}(J)\left[(a_\mathbf{p}^s)^\dagger a_\mathbf{p}^s + \Lambda_J^{-1}(a_\mathbf{p}^s)^\dagger \Lambda_J^{-1}a_\mathbf{p}^s\right] \quad (54)$$

where $E_\mathbf{p}(J)$ is the energy of the Majorana state with spin $J$. Hamiltonian **(54)** is lower bound by the tachyonic term that, therefore, corresponds to the contribution given by the antiparticle in the Dirac Hamiltonian. The particle energy always remains positive and tends to zero, increasing the number of tachyons *bound* to the bradyon with $J = 1/2$. Therefore, the Majorana particle with positive energy can never be superluminal because of the antitachyon that decreases the energy as the particle velocity approaches the speed of light, which is similar to the process that occurs in the Dirac theory when the antiparticle is considered. If the theory had been constructed using the tachyonic field with positive energy, then the Hamiltonian would never been lower bound and would have suffered the same troubles encountered when the quantisation of Majorana field is faced directly [12]. Moreover, since the energy of the bradyon is lower bound (never lower than zero), the energy of the antitachyon cannot be infinite or zero. This explains why the Majorana *exotic states* are made by a bradyon particle and antitachyon particles.

Regarding the Majorana field, since the Lagrangian is given by the contribution of the Dirac field with positive energy and the tachyonic field with negative energy, the following equation can be obtained:

$$J_M^\mu(x) = \bar{\psi}_M \gamma_M^\mu \psi_M = \psi_{D_+}^\dagger \psi_{D_+} + \psi_{D_+}^\dagger \gamma^\mu \psi_{D_+} + \psi_{L_-}^\dagger \psi_{L_-} + \psi_{L_-}^\dagger \gamma^\mu \psi_{L_-} =$$
$$= \psi_{D_+}^\dagger \psi_{D_+} + \psi_{D_+}^\dagger \gamma^\mu \psi_{D_+} + \Lambda_J^{-1} \psi_{D_+}^\dagger \gamma^\mu \Lambda_J^{-1} \psi_{D_+} \quad \gamma = 1,2,3 \quad (55)$$

The current is given by a time component and two spacelike components, of which one corresponds to the bradyonic aspect and the other corresponds to the tachyonic aspect. The divergences of the current **(55)** and any of its components are zero.

Lastly, in the Majorana field, the vacuum energy corresponds to an *exotic state* with infinite value of the spin. Therefore, the Majorana quantum vacuum may form particle-antitachyon pairs, and a luxon is a *composite* particle in the limit of infinite $J$. The idea of luxon being a composite particle originates from the beginning of quantum mechanics and has been reconsidered in the last decades in the ambit of the high energy physics [23-24].

## 7   Particle-Tachyon Interaction Terms

In addition, there is an interaction term between a bradyon and a tachyon that responsible for the existence of the Majorana *exotic states*. Since no experimental data exists corresponding to this interaction, its physical nature can be determined based on the results of previous studies and some inherent properties of the theory. The theoretical assumptions are as follows:

1) The kinematic theory of Lemke on the decay of an ordinary particle predicts the emission of luxons $\gamma$ [25].
2) The Majorana states with spin $J$ are formed only at very high energies and, considering the uncertainty principle, they have very short lifetimes that are inversely proportional to the spin. Therefore, the interaction force is of a *weak* nature.
3) Considering an electrically charged bradyon, antitachyons forming the composite particle may be charged or neutral. (the Majorana equation applies both for charged and neutral particles, as well as luxons). Therefore, we may assume that the tachyonic tower also has an electric charge (which is indicated with $q_t$) that interacts with the electric charge of the bradyon.

Thus, the interaction between bradyon and antitachyons always has an exchange component (the STL matrices are just the algebraic formulation of this interaction) and if the bradyon is in its fundamental state, the interaction has a component of *electrical* nature. Regarding the first interaction term, the following Fermi method can calculate the weak interaction [26-27]:

$$\mathcal{L}_{int.}(exchange) = G_F(\bar{\psi}_{L_-} \gamma^\mu \psi_{D_+}) + c.c. = G_F(\Lambda_J^{-1} \psi^\dagger \gamma^\mu \psi) + c.c. \quad (56)$$

where c.c. represents complex conjugate. Regarding the interaction between Coulomb charges and tachyonic charges, their respective vector potentials can be calculated by the following:

$$\mathcal{L}_{int.}(charge) = i[qA_\mu \bar{\psi}_{D_+} \gamma^\mu \psi_{D_+}]i[q_t A_{\mu_t} \bar{\psi}_{L_-} \gamma^\mu \psi_{L_-}] \quad (57)$$

where $A_\mu$ is the vector potential of the electromagnetic field and $A_{\mu_t}$ is the potential vector of the tachyonic field. The latter is potentially connected to the *fifth force* [28], and the source transporting the tachyonic and electric charge always has subluminal behaviour. In addition, the potential generated by the tachyonic charge is not related to the propagation of the tachyonic wave [29]; therefore, it is consistent with the theory constructed since it is inherent to the only bradyonic tower.

The potential vector of the quantum electrodynamics is as follows:

$$A_\mu = \sum_{k,\lambda} \sqrt{\frac{\hbar}{2\omega_k \varepsilon_0}} \left( \varepsilon_{\mu\lambda} \alpha_{k\lambda} e^{ik \cdot x} + \varepsilon_{\mu\lambda}^* \alpha_{k\lambda}^\dagger e^{-ik \cdot x} \right) \quad (58)$$

where $\varepsilon_{\mu\lambda}$ ($\lambda=1,2$) is the polarisation vector of photon while $\alpha_{k\lambda}$ and $\alpha_{k\lambda}^{\dagger}$ are their creator and annihilator operators. The potential vector of tachyons (gauge invariant) can also have the same structure of **(58)**:

$$A_{\mu_t} = \sum_{k,\lambda} \sqrt{\frac{\hbar}{2\omega_{k_t}\varepsilon_0}} \left( \varepsilon_{\mu\lambda_t} \alpha_{k\lambda_t} e^{ik\cdot x} + \varepsilon_{\mu\lambda_t}^* \alpha_{k\lambda_t}^{\dagger} e^{-ik\cdot x} \right) \tag{59}$$

where $\omega_{k_t}$ is the frequency of the luxon mediating the tachyon-tachyon interaction, $\varepsilon_{\mu\lambda_t}$ is its polarisation vector (and therefore is a boson with spin 1), and $\alpha_{k\lambda_t}$ and $\alpha_{k\lambda_t}^{\dagger}$ are their creator and annihilator operators.

Since the Majorana *exotic states* are formed by a half-integer spin bradyon and antitachyons with negative half-integer spin, in the case of a charged particle, the Lagrangian must also include a term describing the field of luxons mediator of the interacting force due to the charge $q_t$. By analogy with the quantum electrodynamics, this field is given by the following:

$$\mathcal{L}_{luxon} = -\frac{1}{4\mu_0} F_{\mu\nu_t} F_t^{\mu\nu} \tag{60}$$

where $F_{\mu\nu_t} = (\partial_\mu A_{\nu_t} - \partial_\nu A_{\mu_t})$. The Lemke Lagrangian can be corrected by introducing an interaction term. Using the method of minimal substitution, the following is obtained:

$$\mathcal{L}_{L_-} = \sum_J \frac{g(\varepsilon(J))}{\sqrt{\varepsilon(J)}} \left[ \Lambda_J^{-1} \psi^\dagger \left( i\gamma^\mu (\partial_\mu - iq_t A_{\mu_t}) \right) \Lambda_J^{-1} \psi - \varepsilon(J) m_0 \Lambda_J^{-1} \psi^\dagger \Lambda_J^{-1} \psi \right] \tag{61}$$

The explicit form of Lagrangian **(1)** is then the following:

$$\left\{ \sum_J \frac{g(\varepsilon(J))}{\sqrt{\varepsilon(J)}} \left[ \Lambda_J^{-1} \psi^\dagger \left( i\gamma^\mu(\partial_\mu - iq_t A_{\mu_t}) \right) \Lambda_J^{-1} \psi \right] + \psi^\dagger (i\gamma^\mu \partial_\mu) \psi - m_0 \psi^\dagger \psi - \frac{1}{4\mu_0} F_{\mu\nu_t} F_t^{\mu\nu} \right\} =$$

$$= G_F \left[ (\Lambda_J^{-1})^T \psi^\dagger \gamma^\mu \psi \right] + \left\{ G_F \left[ (\Lambda_J^{-1})^T \psi^\dagger \gamma^\mu \psi \right] \right\}^* - \left[ q A_\mu \psi^\dagger \gamma^\mu \psi \right] \left[ q_t A_{\mu_t} (\Lambda_J^{-1})^T \psi^\dagger \gamma^\mu \Lambda_J^{-1} \psi \right] \tag{62}$$

## 8 Discussion

This study has shown that the Majorana equation for particle with arbitrary spin *hides* a deeper structure where the particles with increasing spin are composite systems formed by a half-integer spin bradyon and half-integer spin antitachyons. Within the quantum field theory, these systems are produced by the interaction between the Dirac field with positive energies and the tachyonic field with negative energies. This interaction leads to a lower bound Hamiltonian constructed by field operators that are coherent with the spin-statistic theorem. The Majorana field describes these systems by an infinite sum of Dirac fields which differ in mass and the SLT matrix that transforms the spinor of the bradyon in its fundamental state in a tachyon spinor. This approach solves all troubles encountered by directly facing the quantisation of a field with infinite components with positive energies.

The interaction between the Dirac field with positive energies and the tachyonic field leads to interaction terms between bradyon-tachyon, tachyon-tachyon and the respective charges. In addition, the probability of existence of an *exotic state* is proportional to the relativistic factor $\beta$ and, once the energy is established, the probability decreases as spin $J$ increases [22]. This suggests that the *exotic states* are instable and that their instability increases as the number of antitachyons forming the states increases. Based on these argumentations, the bradyon-tachyon interaction is a *weak* nature. Furthermore, the previous section proved that this interaction is given by the transformation of the bradyon spinor performed by the SLT matrices. The SLT matrices depend on the spin value $J$, but the explicit form of this algebraic dependence has not yet been determined. To do this, the numerical factor $\eta(J) = \sqrt{(J + 1/2)^2 - 1}/(J + 1/2)$ that ranges between zero, when the bradyon is in its fundamental state corresponding to $J = 1/2$ (minimum energy), and one, when the bradyon is in an *exited state* with a spin value that tends to infinite (maximum energy corresponding at the limit $v \to c$), must be determined. Therefore, the factor $\eta(J)$ coincides with the relativistic factor $\beta$, and since the SLT matrices are constructed using this factor [30], their functional dependence becomes explicit.

Overall, the Majorana quantum field theory is formulated simply by Dirac gamma matrices and SLT matrices, both of finite dimensions, which act on a Dirac four-spinor. However, the nature of the bradyon-tachyon interaction and, above all, the tachyon-tachyon interaction must be investigated, not only for the purely quantum aspect, but also for aspects concerning their charge (not necessarily of electrical nature). This information does not directly emerge from the Majorana equation and goes beyond the main purpose of this study, which aims is to find an alternative way to quantise an infinite component field that is consistent with the laws of quantum theory. Even regarding the fine structure of the internal *exotic states*, due to the projection of the total spin about the z-axis, a more in-depth analysis is needed. Consequently, the investigation of these aspects will be postponed to a more specific study.

## 9   Conclusion

The quantisation of Majorana field allows the study of new composite quantum systems whose stability is achieved only in extreme energy conditions where the bradyonic field interacts with the tachyonic field [31]. The internal structure of these *exotic* systems can be imagined as a planet (the bradyon) orbiting a system of stars (an even number of antitachyons) such that the total spin is $n/2$ (where $n$ is an integer number) and the mass is always real. These systems may be the precursors or a part of the *primordial particle broth* which gave rise to the topical quantum particles. High energy composite particles of *exotic* nature have already been detected in LHCb [32], and experimentalists are engaged in the research of similar *objects* that better explain the matter genesis (in all its forms). The results obtained in this work also show that elementary particles have an internal structure, at least under extreme conditions, which could be the starting point for explaining phenomena not yet well understood, such as the oscillation of the particle mass [33] or the lack of fermions with high spin values, and physicists are trying to explain these concepts by string theory. In this sense, the Majorana field theory could be the *link* between the quantum mechanics and the new theoretical models where elementary particles are *objects* with spatial extensions. A proof of this *link* is that the spin of a hadron is never greater than a certain multiple of the root of its energy. No simple hadronic model, such as the model that considers particles as composed of a set of smaller particles interacting by a force, explains these relationships [34]. Using the energy-momentum relation for whatever Majorana *exotic state*, it is easy to find that the total spin $J$ is given by the following:

$$J = \sqrt{E}\,\frac{2\gamma E_0 - E}{E\sqrt{E}} \qquad (63)$$

where $E$ is the relativistic energy of the *exotic state*, $E_0$ is the energy of the bradyon in its fundamental state and $\gamma$ is the Lorentz factor. Equation **(63)** proves that the spin of a particle is a certain multiple of the root of its energy, confirming the above statement.